# Design and Characterization of High Efficiency Single-stage Electromagnetic Coilguns

**Sophia Chen, Annie Peng, Ava Chen, and Takyiu Liu**
West Valley College, Saratoga, CA 95070 USA
Corresponding author: Takyiu Liu (e-mail: takyiu.liu@westvalley.edu).

**ABSTRACT** This study presents several novel approaches to improve the efficiency of a single-stage coilgun. Conventional designs typically feature a uniformly wound solenoid and a ferrite projectile. For our research, we constructed a microcontroller-based prototype to test several new enhancements including the use of a bipolar current pulse, a stepped multilayer coil with non-uniform winding densities, and the replacement of conventional ferrite projectiles with a neodymium permanent magnet. These modifications were designed to reduce energy loss and improve projectile acceleration by changing magnetic field strength and effectively controlling the magnetic flux. The experimental results show that the proposed methods resulted in significant efficiency improvements, with the varying current pulse and stepped coil design providing enhanced magnetic force at key points in the projectile's path, and the permanent magnet projectile contributing to higher velocities and efficiencies by leveraging the current pulses. Our findings suggest that combining these enhancements significantly improve coilgun performance, achieving higher velocities and efficiencies. These findings can be applied to future coilgun developments, such as multi-stage coilgun systems.

## I. INTRODUCTION

Electromagnetic launchers utilize electromagnetic forces to propel projectiles at significantly higher velocities than traditional chemical-based launch systems. These technologies, which have been extensively researched for military and transportation applications, are generally categorized into two types: railguns and coilguns.

Railguns generate Lorentz forces by driving extremely high currents (in the kilo- to mega-Ampere range) through a pair of parallel conducting rails and a contacting armature. However, the severe mechanical and thermal stresses experienced by the rails and contacts lead to rapid wear and tear, resulting in high maintenance costs and frequent downtime [1].

In contrast, coilguns employ magnetic forces to accelerate a ferrite or magnetic payload without requiring high currents to pass directly through the projectile or ballast. While coilguns avoid the maintenance challenges faced by railguns, they typically struggle with lower velocities and efficiency. This study investigates various strategies to improve the velocity and efficiency of coilgun designs.

## II. PHYSICS OF A COILGUN

Efficiency of the coilgun is calculated as the conversion of electrical energy to mechanical energy:

$$\eta = \frac{\frac{1}{2}mv^2}{\frac{1}{2}CV_i^2 - \frac{1}{2}CV_f^2}$$

where $m$ = mass of the projectile, $v$ is the exit velocity, $C$ is the capacitance of the capacitor, and $V_i$ and $V_f$ are the initial and final voltages of the supply capacitor.

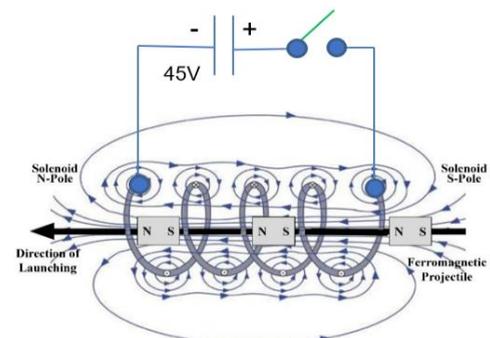

**Figure 1. Coilgun Solenoid**

Coilguns operate by sending a current pulse through a coil, generating a transient magnetic field that exerts force on a projectile. Higher launch velocities and efficiencies can be achieved by adjusting the current pulse, optimizing the coil winding, and selecting the appropriate projectile type. Refer to Fig. 1 as we review the underlying physics.

### 1) ELECTROMAGNETIC INDUCTION

When an electric current flows through a coil, it generates a magnetic field $\vec{B}$. The strength and direction of this field are determined by the current and the number of turns in the coil. Like Slade [2], we ignore eddy currents and edge effects, the latter because of "long-coil" geometry where coil length is much greater than radius, the former when using sintered or laminate projectiles.

### 2) MAGNETIC FORCE ON THE FERRITE PROJECTILE

Ferrite is a type of ceramic compound of iron oxide ($Fe_2O_3$) with a high magnetic permeability. Before the coil is energized, the ferrite rod is in its natural, unmagnetized state. The ferrite rod is momentarily magnetized when the magnetic field penetrates into the rod, effectively turning it into a magnet. Let's denote the resultant magnetic dipole moment $\vec{m}$.

The potential energy of a magnetic dipole $\vec{m}$ in a magnetic field $\vec{B}$ is given by the dot product: $U = -\vec{m} \cdot \vec{B}$. The magnetic force $\vec{F}$ is the negative gradient of this energy:

$$\vec{F} = -\nabla U = \nabla(\vec{m} \cdot \vec{B}) = \left( \frac{\partial(\vec{m}\cdot\vec{B})}{\partial x}, \frac{\partial(\vec{m}\cdot\vec{B})}{\partial y}, \frac{\partial(\vec{m}\cdot\vec{B})}{\partial z} \right).$$

### 3) LOSS OF INDUCTION

Once the ferrite rod leaves the influence of the coil, the magnetic field is no longer present to keep the magnetic domains of the rod aligned. The rod then loses its magnetization and returns to its unmagnetized state.

## III. DESIGN OPTIMIZATIONS

Since the force acting on the ferrite rod is proportional to the gradient of the magnetic field, the rod accelerates toward the region where the magnetic field is strongest. In a uniform coil, however, the magnetic field is nearly constant inside, contributing little to the coilgun's velocity and efficiency. The most favorable outcome in this scenario is that the ferrite rod is simply drawn to the center of the uniform coil. If the current remains active as the rod moves from the center to the end of the coil, the ferrite will experience a decelerating 'suckback' force. Therefore, precise timing is essential to switch the coil current on and off at the optimal moments to achieve maximum efficiency.

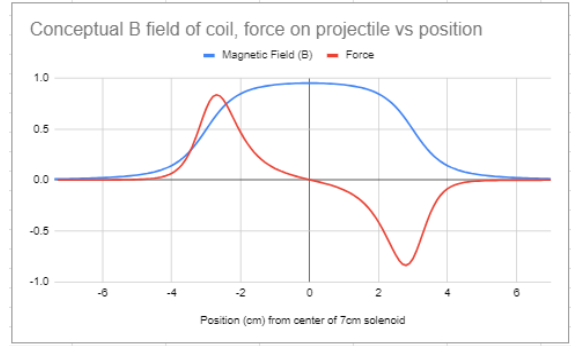

Figure 2. Magnetic field and suckback force of typical coilgun

The use of a ferrite rod presents two key challenges:
1) The rod's magnetization can become momentarily saturated, limiting its effectiveness.
2) Even if the coil current is reversed, the rod's magnetization will also reverse, causing it to continue experiencing a decelerating magnetic force as it moves from the center of the coil to the end.

Both issues can be resolved by switching to a permanently magnetized projectile. A permanent magnet will experience both attractive and repulsive forces as it moves through the coil if the current is reversed during its traversal.

Typical coilgun designs in the literature [2] [3] [5]:
1) Launch a ferrite rod as the projectile.
2) Use a coil with uniform windings.
3) Rely on dumping a large amount of current into the coil from a capacitor.

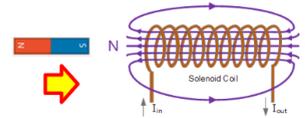

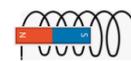

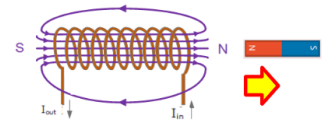

Figure 3. Using magnet projectile and bidirectional current to avoid suckback

Our design focuses on optimizing the coilgun in several key areas, including:
1) Use of a permanent N52 sintered neodymium magnet as the projectile.

2) Development of coils with non-uniform windings to leverage $\frac{\partial(\vec{m}\cdot\vec{B})}{\partial x}$
3) Use of a microcontroller to precisely control the on-off switching of the coil currents [4].
4) Incorporation of an H-bridge to reverse the current direction in the coil when the N52 magnet reaches the region of maximum magnetic field. The permanent magnet is initially drawn toward the coil's center, and upon current reversal, it is propelled out of the coil. This reversal actively cancels any residual current, minimizing suckback.
5) Recording the current data onto a microSD card every millisecond for subsequent analysis.
6) We developed a prototype to validate these concepts. The following sections will detail the implementation and present the experimental results.

## IV. DESIGN OF THE OVERALL MASTER CONTROL SYSTEM

The master control system is managed by an Arduino Mega 2560 microcontroller, while Arduino Nano microcontrollers handle the digital leveling and the sensors for measuring velocity and current.

Figure 4 shows the prototype. These are the highlights:

1) Sections 2 and 3 feature standalone photogates controlled by Arduino Nano microcontrollers, mounted on a 3D-printed stage to provide precise velocity measurements as the projectile exits the tube. Each photogate consists of a module that includes a photo interrupter (IR emitter and phototransistor) and an LM393 comparator.
2) Section 6 includes an IMU6050 (3-axis gyroscope) digital bubble level, which ensures the precise horizontal alignment of the tube.
3) Section 8 consists of four solid-state relays (SSRs) configured as an H-bridge.
4) Section 9 are flyback diodes to protect the 4 SSRs from voltage spikes from the coil.
5) Section 10 is a Hall-effect current sensor.
6) Section 11 features a microSD card reader. We utilize hardware interrupts to accurately log all current data every millisecond.
7) Section 12 is the Arduino Mega 2560 microcontroller.
8) Section 14 is a charging/discharging switch.
9) Section 15 is our current source for charging the capacitor: a 50V, 0.12F power supply.
10) Section 17 is our charging power supply.
11) Section 18 is the power source for digital controls.

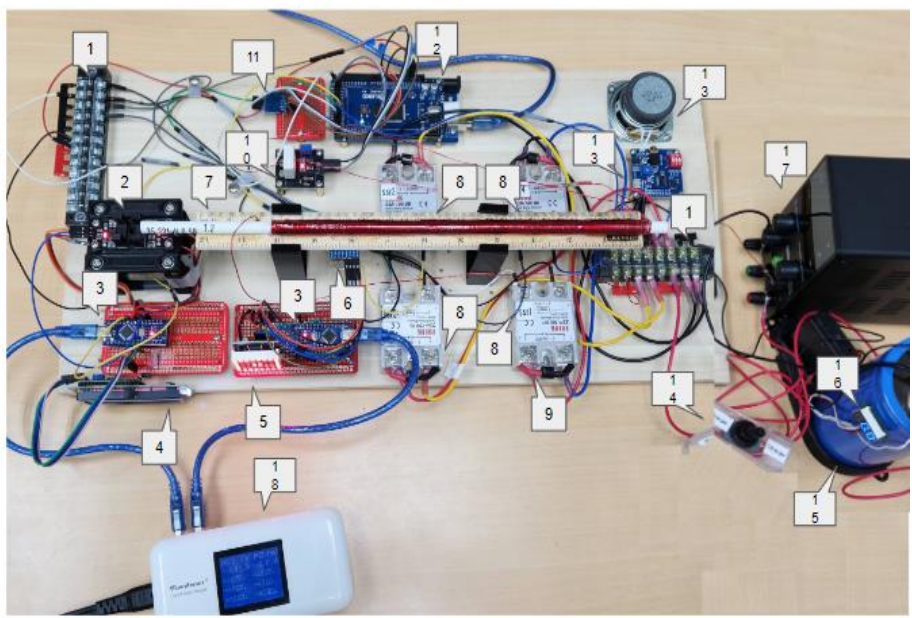

### General Instrumentation

1. Terminal Blocks
2. LM393-photogates on 3D printed stage for precise speed measurements
3. Stand-alone Arduino Nano controllers for speed measurement
4. LCD display for speed and time interval
5. LED matrix for IMU 6050
6. IMU 6050 for digital bubble
7. Coil (standard coil) under test
8. Solid state relays (4) 100V/100A
9. Flyback diodes (one labeled)
10. 100A Current sensor
11. Arduino SD card data logger
12. Arduino Mega - master controller
13. Arduino MP3/speaker for launch announcement
14. Charging/discharging switch
15. 45V, 0.12F capacitor
16. Voltmeter
17. Charging Power supply
18. Power source for stand-alone microcontrollers

**Figure 4. Experimental Apparatus**

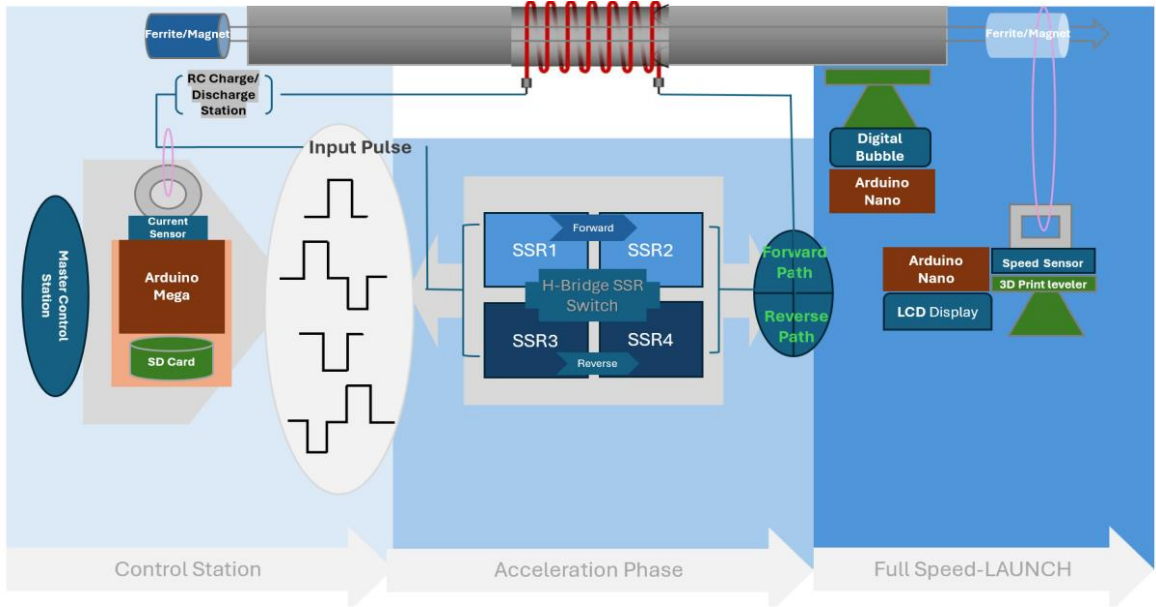

**Figure 5. Master Control Station**

### A. MASTER CONTROL BOARD

Our Master Control Board utilizes an Arduino Mega 2560. It generates a 5V digital signal to control four SSR (100V, 100A DC) relays, triggering them ON and OFF in pairs to manage the direction and duration of the current pulse in the solenoid coil. The board measures the output voltage from the current sensor and records the data onto an SD card using timer interrupts.

The gyroscope and speed measurement systems are on separate boards. The speed measurement system employs two photogate sensors, positioned a fixed distance apart, which use hardware interrupts to accurately timestamp sensor changes and calculate the projectile's speed at the tube's exit. The gyroscope (IMU 6050) displays its data on an 8x8 LED matrix, which is cleared after each measurement to provide real-time data.

### B. CURRENT LOGGING & SD LOGGER

A Hall effect sensor (up to 100A, EC) measures the current flowing through the coil. A microSD Card Module with a 2GB microSD card, using an SPI interface, is employed to store the data collected from the current sensor. The change in current for our standard coil is measured at 0.03 volts per ampere.

### C. DESIGN OF THE CURRENT REVERSAL SYSTEM

We designed and constructed custom H-bridges to maximize current handling capability. Each H-bridge comprises four single-phase solid-state relay (SSR 100DD) modules. These SSR modules accept input voltages ranging from 3-32V and can handle power side voltages from 5-200V DC, with a current rating of up to 100A. We selected 100A/100V solid-state relays for their sub-millisecond switching speed, in contrast to mechanical relays, which have switching times of approximately 10ms. To initiate the current in the coil, the Arduino activates SSR1 and SSR2 while keeping SSR3 and SSR4 deactivated. To reverse the current flow when the permanent N52 magnet reaches the position of maximum magnetic field, SSR1 and SSR2 are turned off and SSR3 and SSR4 are turned on. An illustration of this process is shown in Fig. 6.

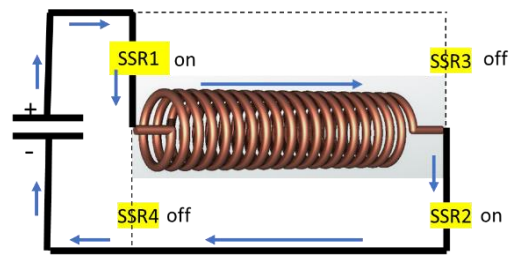

**Figure 6. H-bridge with SSR1 and SSR2 on**

When switching approximately 50A in ~0.1ms through an inductor of around 100µH, around 50V of back EMF is generated. To protect the SSRs from potential damage caused by back EMF and voltage spikes, flyback diodes are used in parallel with each SSR. These diodes are reverse-biased during normal operation and do not affect the H-bridge's performance. When a voltage spike occurs, the flyback diodes become forward-biased and conduct, safely directing the charge into the capacitor and maintaining energy efficiency.

In conventional coilguns, energy is typically dissipated through diodes (shunted with the coil) and resistors rather than being recycled into the capacitor.

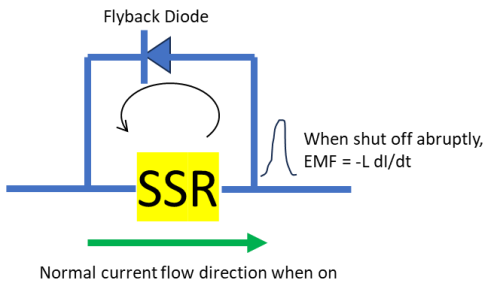

Figure 7. Back EMF Mitigation

The SSR input array is user-configurable, allowing for the storage of SSR states and durations to customize buffer time and current direction in the circuit. Timer interrupts are utilized to prompt the Arduino Mega to log voltage values with timestamps onto the SD card every millisecond. Once the array is full, the data is written to the microSD card.

A waveform from one run is shown in Figure 8. The F5 B5 R5 B10 F5 B5 R4 tag refers to the sequence of forward/buffer/reverse pulse duration in milliseconds for that experimental run.

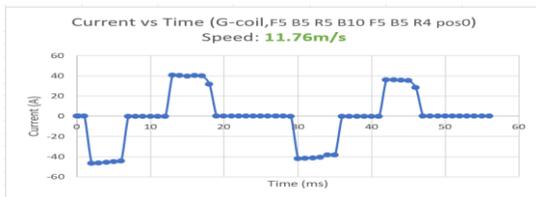

Figure 8. Measured Current Pulse

*D. COIL TRANSIENT SIMULATION*

Before hand-winding the coils, we conducted circuit simulations to validate the coil design parameters and to compare against actual current measurements. We performed LTspice® simulations using a lumped LR model for the target coil and a 0.12F capacitor with a much smaller equivalent series resistance compared to the coil. For simplicity, we omitted the SSR relays and flyback diodes, focusing solely on the coil. Based on these simulations, we selected the target coil parameters as follows: inductance L=36 µH and resistance R=0.6 Ω.

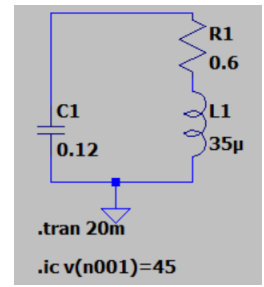

Figure 9. Lumped Model of Capacitor Discharge into Solenoid

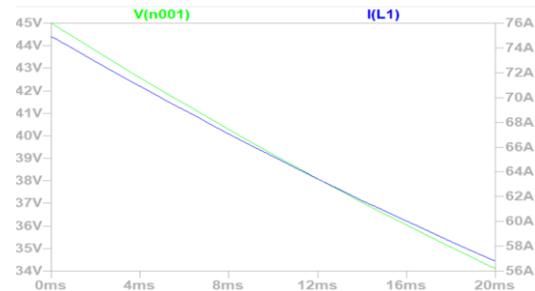

Figure 10.
SPICE Simulation to Select Coil Parameters

## V. DESIGN OF THE COILS

The solenoid coils are constructed using 10 meters of 22 AWG enameled copper wire with a diameter of 0.0256 inches. The coils are wound onto the surface of 14-inch polystyrene tubes, each with an outer diameter of 7.9 mm.

We characterized an assortment of winding profiles as illustrated below:

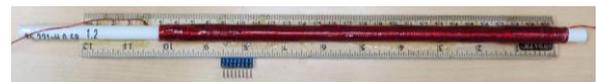

Figure 11. Single-Layer Coil

The single-layer coil has an inductance of 34.2 µH and a resistance of 0.5 Ω. This configuration represents standard coilguns used in industrial applications, providing a baseline for comparing performance with other modified coil designs. It is also used to evaluate the effects of different payloads, such as ferrite rods versus permanent neodymium N52 magnets.

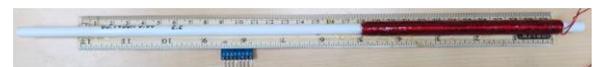

Figure 12. Double-Layer Coil

The double-layer coil is characterized by an inductance of 61.9 µH and a resistance of 0.6 Ω. With two layers of uniform winding, this configuration enhances the magnetic flux concentration, thereby increasing the force exerted on the projectile. The improved magnetic field contributes to better efficiency compared to the single-layer coil.

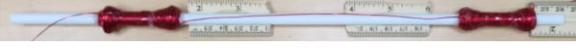

**Figure 13: Dual 9-5-1-5-9 Coil (Coil G)**

The dual 9-5-1-5-9 coil has an inductance of 94.9μH and a resistance of 0.6Ω. This non-uniform design simulates a rapidly changing coil configuration with six equal sections on each of the two smaller coils, which are separated by a gap. The six sections have nine coils, five coils, one coil, one coil, five coils, and nine coils, respectively. The nine-layer sections are positioned at the start and end of each smaller coil to produce a stronger magnetic field and achieve greater acceleration at the entry and exits.

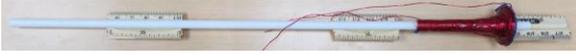

**Figure 14. Exponential Coil**

The exponential coil has an inductance of 127.6 μH and a resistance of 0.7 Ω. This design models an exponential function, where the number of layers decreases progressively along the length of the coil. The intense magnetic field at the start creates a powerful initial acceleration.

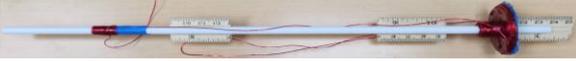

**Figure 15. T-Shaped Coil**

The T-shaped coil is characterized by an inductance of 147.4μH and a resistance of 0.6Ω. It has a section of 30-layer coiling, and a section of four-layer coiling; there is also a small section of single-layer coil at the end. This design concentrates the magnetic field at the entry while maintaining a relatively strong magnetic field (four layers) on the rest of the coil.

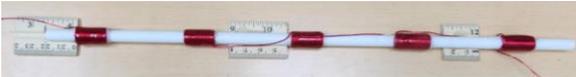

**Figure 16. Linear Accelerator Coil**

The linear accelerator coil has an inductance of 49.5μH and a resistance of 0.7Ω. This design reaccelerates the projectile multiple times while it is moving in the tube by using five equal sections of double-layer coiling.

## VI. MAGNETOSTATICS OF SELECT COIL WINDINGS

The magnetic field in the first few configurations can be estimated by modeling the discrete windings as continuous infinitesimal coils carrying uniform current. First, we consider a single loop. The radial field cancels at the center axis of the solenoid, so we only need to consider the axial field which can be calculated using Biot-Savart law applied on the loop:

$$B_{loop} = \frac{\mu_0 I}{4\pi r^2} \oint ds = \frac{\mu_0 I}{4\pi r^2} \int_0^{2\pi} \frac{R^2 d\theta}{r} = \frac{\mu_0 I}{4\pi r^2} \frac{2\pi R^2}{r}$$

$$= \frac{\mu_0 I R^2}{2(x^2 + R^2)^{\frac{3}{2}}}$$

where $\mu_0$ is the permeability of space, $I$ is the current, $r = \sqrt{x^2 + R^2}$ is the distance to the origin from an infinitesimal arc of one loop of radius $R$ at location $x$, and $ds = |d\vec{s} \times \hat{r}| = \frac{R}{r}|d\vec{s}| = \frac{R}{r}(Rd\theta) = \frac{R^2 d\theta}{r}$ is the axial projection of the field. To get the total magnetic field of one layer of $N$ windings with current density $\frac{NI_0}{L}$, we integrate the contributions along the length $L$ of the solenoid whose leftmost edge is at $L_0$ and use $dI = \frac{NI_0}{L}dx$:

$$B_{single} = \frac{\mu_0}{2}\frac{NI_0}{L}\int_{L_0}^{L_0+L} \frac{dx}{(x^2+R^2)^{3/2}}$$

$$= \frac{\mu_0}{2}\frac{NI_0}{L}\left(\frac{L_0+L}{\sqrt{(L_0+L)^2+R^2}} - \frac{L_0}{\sqrt{L_0^2+R^2}}\right)$$

For the double layer coil, the first set of windings has radius $R$ and second set has radius $R+d$ where $d$ is the incremental distance, which is 0.0256 inches for 22 AWG winding wire:

$$B_{dual} = \frac{\mu_0}{2}\frac{NI_0}{L}\left(\frac{L_0+L}{\sqrt{(L_0+L)^2+R^2}} - \frac{L_0}{\sqrt{L_0^2+R^2}}\right.$$
$$+ \frac{L_0+L}{\sqrt{(L_0+L)^2+(R+d)^2}}$$
$$\left. - \frac{L_0}{\sqrt{L_0^2+(R+d)^2}}\right)$$

For the specific case of uniform multilayer windings from $R_o$ outer radius to $R_i$ inner radius with current density $\frac{N}{L}\frac{I_0}{R_o-R_i}$, we integrate the contributions across the radius from $R_o$ to $R_i$ to get:

$$B = \frac{\mu_0}{2}\frac{N}{L}\frac{I_0}{R_o-R_i}\int_{R_i}^{R_o}\left(\frac{L_0+L}{\sqrt{(L_0+L)^2+R^2}}\right.$$
$$\left. - \frac{L_0}{\sqrt{L_0^2+R^2}}\right)dR$$

$$B = \frac{\mu_0 N I_0}{2L(R_o-R_i)}\left((L_0+L)ln\left(\frac{\sqrt{(L_0+L)^2+R_o^2}+R_o}{\sqrt{(L_0+L)^2+R_i^2}+R_i}\right)\right.$$
$$\left. - L_0 ln\left(\frac{\sqrt{L_0^2+R_o^2}+R_o}{\sqrt{L_0^2+R_i^2}+R_i}\right)\right)$$

We can use these quasi-analytic magnetostatic formulas to compare against numerical simulations and measured observations.

## VII. COMPUTER SIMULATION

We can gain valuable insights through computer simulation. The total magnetic field at any point along the tube can be modeled using a Python program. In our simulation, we use the wire width (0.68 mm) as the fundamental unit of length, digitizing the 14-inch tube into 523 current loops. The magnetic field produced by a single current loop can be analytically calculated at any point along the tube's axis. By superimposing the magnetic fields from all the loops, we can determine the overall magnetic field distribution. Additionally, analyzing the magnetic field gradients between adjacent loops allows us to derive the magnetic force on a magnetic dipole (our permanent magnet payload). Below are the results from some of the simulations.

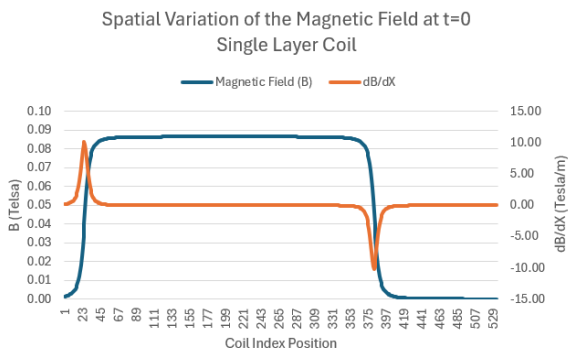

**Figure 17(a).** Magnetic field B, dB/dx of a single-layer coil at t=0

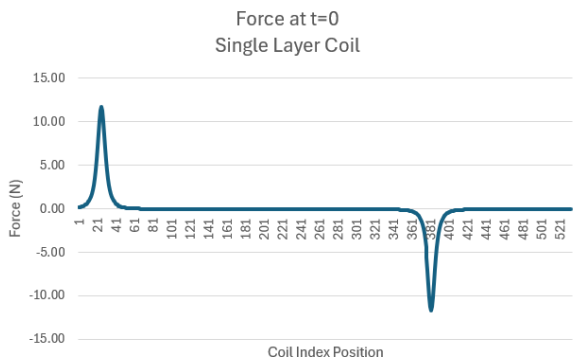

**Figure 17(b).** Force of a single-layer coil at t=0ms

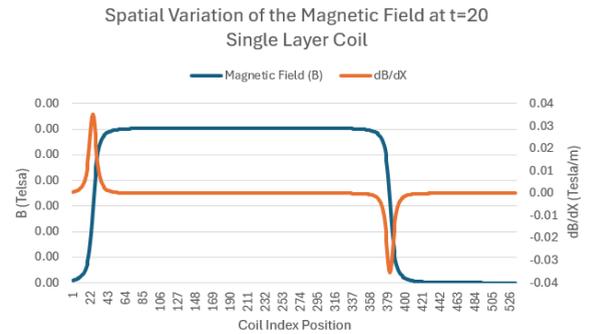

**Figure 18(a).** Magnetic field B, dB/dx of a single-layer coil at t=20

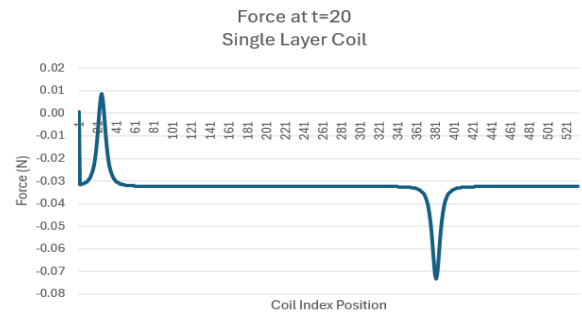

**Figure 18(b).** Force of a single-layer coil at t=20ms

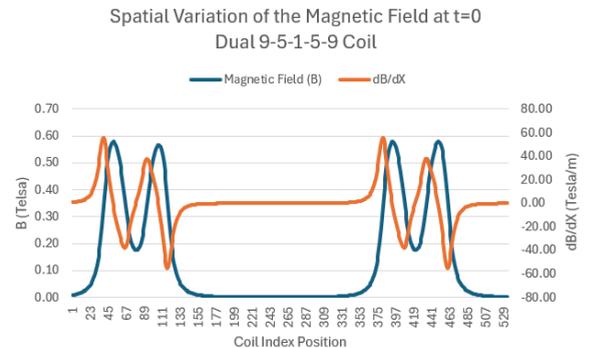

**Figure 19(a).** Magnetic field B, dB/dx of a dual 9-5-1-5-9 coil at t=0

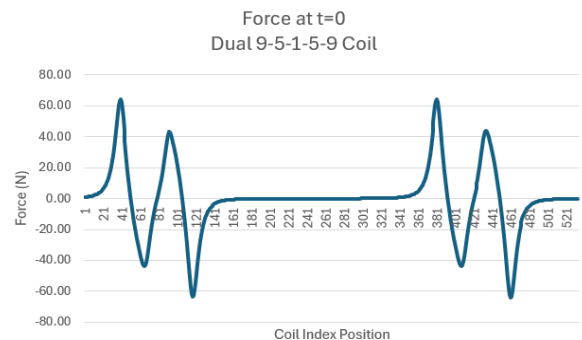

**Figure 19(b).** Force of a dual 9-5-1-5-9 coil at t=0

## VIII. EXPERIMENTAL DETERMINATION OF OPTIMAL INITIAL PLACEMENT

We determined the optimal initial placement experimentally for each coil and measurement run. The following charts display measurements for both ferrite and N52 projectiles with the single-layer coil, relative to various displacements at the entrance of the solenoid barrel. The N52 projectile demonstrates higher efficiency and launch velocity, requiring a smaller displacement to achieve optimal performance.

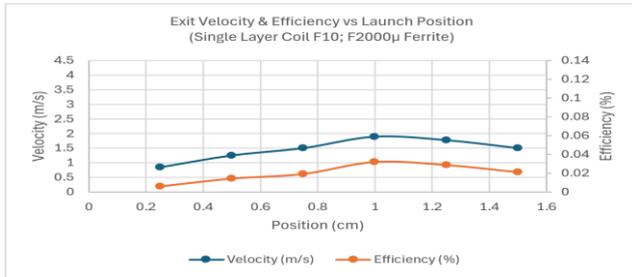

**Figure 20.** Velocity and efficiency vs initial displacement for ferrite projectile and single-layer coil; current pulse: F10 forward 10ms

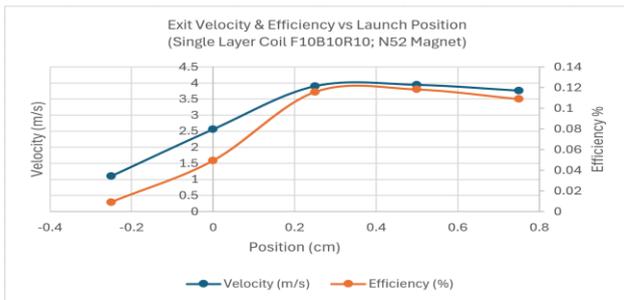

**Figure 21.** Velocity and efficiency vs initial displacement for N52 magnet projectile and single-layer coil; current pulse: Forward 10ms Baseline 10ms Reverse 10ms

The optimal displacement varies with the winding profile. The charts for the G Coil (9-5-1-5-9 layers) and the T Coil are shown in the following figure:

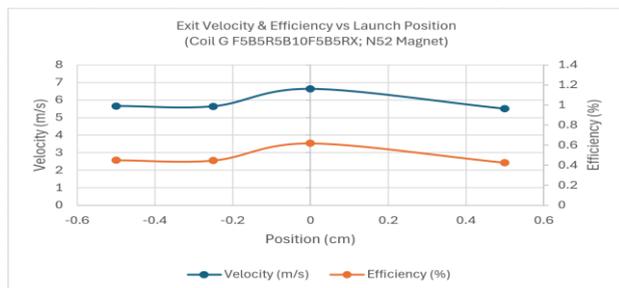

**Figure 22.** Velocity and efficiency vs initial displacement for N52 magnet projectile and coil G (9-5-1-5-9 layers); current pulse F5B5R5B10F5B5R5

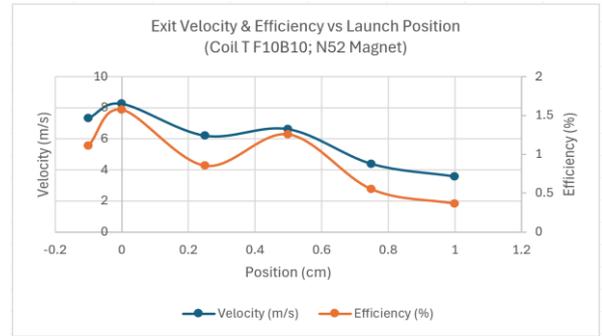

**Figure 23.** Velocity and efficiency vs initial displacement for N52 magnet projectile and coil T; current pulse: forward 10ms

## IX. EXPERIMENT SUMMARY

The following table summarizes the maximum velocity and maximum efficiency for each coil. A variety of current pulses were tested; the launch profile resulting in the highest measurement is recorded in the table using the notation explained earlier: For example, the F6 B5 R6 tag refers to the sequence of 6ms forward pulse, 5ms buffer, and 6ms reverse current pulse for the experimental run yielding the 1.6% efficiency for the 9-5-1-5-9 G coil, the second highest efficiency observed. The T coil had the highest efficiency 1.78% as expected since G and T maximize the gradient $\frac{\partial(\vec{m} \cdot \vec{B})}{\partial x}$ via their winding profiles. The dual G resembles a multistage coilgun and achieves the highest velocity but sacrifices efficiency.

**TABLE I**
**Velocity and Efficiency of Tested Configurations**

| Coil ID | Max Velocity, Current Profile | Max Efficiency, Current Profile [a] |
|---|---|---|
| Standard (ferrite) | 2.64m/s, F50 | 0.03%, F10 |
| Standard (single, magnet) | 4.32m/s, F32 | 0.19%, F14 |
| Standard (magnet) | 6.34m/s, F18 B41 R18 | 0.23%, F14 B45 R14 |
| Double (magnet) | 8.79m/s, F14 B6 R14 | 0.53%, F5 |
| Three Discs (Coil 3M) | 3.26m/s, F10 | 0.14%, F10 |
| Linear Accelerator | 11.27m/s, F5 B5 F5 B4 F4 B3 F2 B1 F2 | 0.86%, F5 B5 F5 B4 F4 B3 F2 |
| Dual 9-5-1-5-9 (Coil G) | 11.76m/s, F5 B5 R5 B10 F5 B5 R4 | 1.11%, F5 B5 R4 |
| Exponential (Coil S) | 11.03m/s, F9 B5 R5 | 1.49%, F9 |
| T-shape (Coil T) | 8.47m/s, F5 B16 R3 | 1.78%, F5 |

[a] Legend for the Current Profile: (F)orward, (B)uffer, (R)everse current pulses with millisecond duration.

## X. DISCUSSION

Our study shows a significant increase in coilgun efficiency, achieving 1.78%, compared to the typical range of 0.3% to 1% reported in previous literature [6] [7]. This improvement was achieved through multiple strategic modifications.

One critical modification was optimizing the launch position of the projectile. The initial launch position affects the magnetic force and acceleration of the coil. If the projectile is placed at a suboptimal position, it is more likely to miss the peak acceleration point, leading to a lower efficiency and exit speed. By placing the projectile at the optimal position, we ensured the increased magnetic flux interaction at the beginning of the launch would achieve maximum acceleration. Our results show that even minor changes to launch position can significantly improve the speed and efficiency.

Additionally, using non-uniform coil densities improved coilgun efficiency and speed. The varying coil densities allowed a more effective magnetic field profile and generated a higher magnetic flux where needed most. By reversing the current direction at specific intervals, energy loss is minimized, leading to improved speeds and efficiencies.

## XI. SUMMARY AND CONCLUSION

We conducted experiments using a 3.73g ferrite rod and a 6.06g N52 neodymium magnet and compared their performance with a standard single-layer coil. The neodymium magnet consistently outperformed the ferrite rod, achieving significantly higher launch velocities, particularly when a reverse current pulse was applied. Without current reversal, the magnet payload demonstrated nearly 1.6X the velocity and 6X more efficiency compared to the ferrite payload. With current reversal, the magnet payload achieved up to 2.4X the velocity and 7.7X the efficiency. .

A standard single-layer coil configuration with a single pulse and a ferrite rod (2.64 m/s exit velocity, 0.03% efficiency) serves as a reference for all the tested coils. The highest velocity was achieved with the neodymium magnet using a dual 9-5-1-5-9 coil (Coil G), reaching 11.76 m/s with a current profile of F5 B5 R5 B10 F5 B5 R4. This setup represents a 4.5X improvement while maintaining a high efficiency of 1.11%. The best efficiency was achieved with a T-shaped coil (Coil T), which reached 1.78% efficiency with a single forward pulse (F5), representing a 59.3X improvement while still achieving a high velocity of 8.47m/s.

Our experiments confirmed that using a current pulse, rather than a constant current, near the edges significantly enhances the conversion efficiency for both uniform and non-uniform coils. Additionally, non-uniform coil densities allow for shorter current pulses. These factors collectively contribute to substantial improvements in payload velocity and efficiency.

## XII. FUTURE WORK

In future work, we plan to implement a daughterboard for a second acceleration pulse and conduct experiments to determine the optimal pulse duration. We plan to explore multi-stage designs and enhancements to the coil designs, including the T-shaped coil, to increase efficiency or velocity. Future research will focus on further improving efficiency, including the use of system simulation (Matlab/Simulink®), field solvers (Ansys Maxwell®) and artificial intelligence to optimize the current profile.

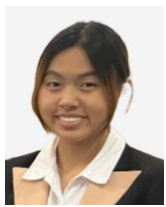

**SOPHIA CHEN** was born in San Jose, California. She is currently enrolled at West Valley College and is a research assistant working under the supervision of Dr. Takyiu Liu. Her interests include robotics and computer engineering.

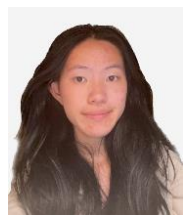

**ANNIE PENG** is a native of San Jose, California. She is presently enrolled at West Valley College and serves as a research assistant under the supervision of Dr. Takyiu Liu.

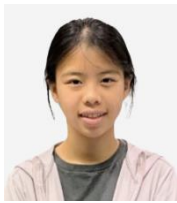 **AVA CHEN** is from San Jose, California. Right now, she's studying at West Valley College and working as a research assistant with Dr. Takyiu Liu.

**TAKYIU LIU** received the B.S. degree in Electrical and Computer Engineering from Hong Kong University in 1984, the M.S. degree in Electrical and Computer Engineering from the University of California, Santa Barbara, in 1986, and the Ph.D. degree in Electrical and Computer Engineering from the University of California, Santa Barbara, in 1990. From 1990 to 1997, he was a Senior Member of the Technical Staff at Hughes Research Laboratories, where he worked on Gas Source Molecular 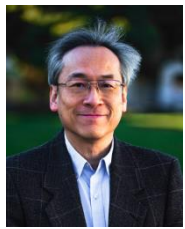 Beam Epitaxy for III-V semiconductors including arsenides and phosphides. He has authored approximately 80 technical papers in semiconductor physics and holds 9 U.S. patents and 3 European patents. Dr. Liu is currently the Chair of the Computer Science and Engineering Department at West Valley College, Saratoga, California. His current research interests include electromagnetic propulsion, the application of machine learning and deep learning in engineering, computer visions and fine-tuning large language models for engineering systems.